\documentclass[intlimits,twoside,a4paper]{article}
\usepackage[cp1251]{inputenc}
\usepackage{bm}
\usepackage[eqsecnum]{cmpj3}

\let\vec\bm

\issue{2020}{23}{4}{43711}
\doinumber{10.5488/CMP.23.43711}

\title[SU(N) to SU(2) symmetry breaking in quantum antiferromagnets]%
{$SU(N)\to SU(2)$ symmetry breaking in quantum antiferromagnets}
\author[A.K. Kolezhuk, T.L. Zavertanyi]{A.K. Kolezhuk\refaddr{KNU,IMAG}, T.L. Zavertanyi\refaddr{KNU}}
\addresses{
\addr{KNU} Institute of High Technologies, Taras Shevchenko National
  University of Kyiv, Ministry of Education and \\Science, 03022 Kyiv, Ukraine
\addr{IMAG} Institute of Magnetism, National Academy of Sciences of Ukraine and
  Ministry of Education and Science of Ukraine, 36--b Vernadskii Av., 03142 Kyiv, Ukraine
}
\date{Received June 22, 2020, in final form August 14, 2020}
\begin{document}

\maketitle

\begin{abstract}
We study a $SU(2)$-symmetric spin-${3}/{2}$ system on a bipartite lattice
close to the antiferromagnetic $SU(4)$-symmetric point, which can be
described by the $CP^{3}$ model with a perturbation breaking the
symmetry from $SU(4)$ down to $SU(2)$ and favoring the N\'eel ordering.  We show that the effective theory
of the perturbed model is not the usual $O(3)$ nonlinear sigma model
(NLSM), but rather the $O(3)\times O(2)$ NLSM.  
We show that in the presence of  perturbation, the topological charge $q$ of the $CP^{3}$ field 
is connected to  the $O(3)$-NLSM type topological charge of the spin texture $Q$ (defined in a usual way via the unit N\'eel vector)
by the relation $q=3Q$,  thus under the influence of the perturbation
unit-charge skyrmions of $CP^{3}$ model bind into triplets. 
We
also show that in the general spin-$S$ case, symmetry breaking from
$SU(2S+1)$ to $SU(2)$ results in the general relation $2S Q_{O(3)}=q_{CP^{2S}}$ between $CP^{2S}$ and $O(3)$ charges, so one can
expect $2S$-multiplet binding of skyrmions.
\end{abstract}

\section{Introduction and model}

Numerous studies in the past two decades have firmly positioned
ultracold gases as an extremely versatile ``toolbox'' capable of
simulating a wide range of problems originating in condensed matter
physics and field theory. Particularly, multicomponent ultracold gases
in optical lattices\cite{Lewenstein+rev07,Bloch+rev08,KawaguchiUeda12} allow one to model spin systems with strong
non-Heisenberg exchange interactions\cite{ZhouSnoek03,ImambekovLukinDemler03,WuHuZhang03,Lecheminant+05,Wu05,Wu06}, normally inaccessible in solid
state magnets. 
 Presence of controllable strong higher-order
(biquadratic, biqubic, etc.) exchange interactions allows one to explore
models with enhanced $SU(N)$ symmetry with $N>2$, which have been a
subject of extensive
theoretical studies
\cite{Affleck86,Affleck88,ReadSachdev89,ReadSachdev90,ItoiKato97,Assaad05,KawashimaTanabe07}.
Realization of  $SU(N)$ antiferromagnets with $N$ up to 10
was suggested\cite{Gorshkov+10,Hermele+09} and realized in experiments \cite{Scazza+14}. Spin systems with strong higher-order exchange interactions may exhibit phases with unconventional (multipole) order and can be considered as a special type of frustrated magnets.

It has been shown \cite{IvKhymKol08,Kolezhuk08} that in
spin-1 systems close to the antiferromagnetic $SU(3)$ point, a perturbation that
breaks this symmetry down to $SU(2)$ can lead to an interesting
effect: unit-charge topological excitations (skyrmions, hedgehogs) of
the effective $CP^{2}$ theory describing the
$SU(3)$-symmetric model bind into doublets that correspond to unit-charge
topological excitations of the effective $O(3)$ nonlinear sigma model
(NLSM) theory describing  the $SU(2)$-symmetric model.

In the present
work, we study spin-${3}/{2}$
systems close to the antiferromagnetic $SU(4)$ point, and show that a
similar effect of binding topological excitations into triplets exists
when the symmetry gets broken down to $SU(2)$.  We further show that
this result can be generalized:
for a system with underlying spin $S$ close to the antiferromagnetic
$SU(2S+1)$ point, a perturbation that brings the symmetry down to
$SU(2)$ can lead to the formation of $2S$-multiplets of topological excitations.

We start with a system of spin-${3}/{2}$ fermions on  a
bipartite optical lattice in $d$ spatial dimensions ($d=1,2$) that can be described by the following
Hamiltonian in the $s$-wave scattering approximation \cite{WuHuZhang03}: 
\begin{equation} 
\label{ham-atom}
\widehat{H}=-t\sum_{\sigma=\pm 1/2,\pm 3/2}\sum_{\langle   ij\rangle} \left(c^{\dag}_{\sigma,i}c^{\vphantom{\dag}}_{\sigma,j} +\text{h.c.}\right)
+\sum_{i}\sum_{F=0,2}U_{F}\sum_{m=-F}^{F} P^{\dag}_{Fm,i}  P^{\vphantom{\dag}}_{Fm,i} \,,
\end{equation}
where $c_{\sigma,i}$ are the spin-${3}/{2}$ fermionic operators at
the lattice site $i$, $t$ is the effective hopping amplitude between
two neighboring sites (which is for simplicity assumed to be the same
for all spatial directions), 
 \[
 P_{Fm,i}=\sum_{\sigma\sigma'}\left\langle
Fm\left|\frac{3}{2} \sigma \right.,\frac{3}{2} \sigma' \right\rangle
c_{\sigma,i}c_{\sigma',i}
\] 
are the operators describing an on-site
pair with the total spin $F$, and interaction strengths 
$U_{0}$, $U_{2}$ are proportional to the scattering lengths in the
$F=0$ and $F=2$ channels, respectively. 

At quarter filling (one particle per site), and in the limit of strong
on-site repulsion $U_{0},U_{2}\gg t$, the charge degrees of freedom
are strongly gapped, and at low excitation energies the system can be described by
the effective Hamiltonian involving only spin degrees of freedom. The
antiferromagnetic $SU(4)$-symmetric point corresponds\cite{KV11} to the limit
$U_{2}\to\infty$ and its Hamiltonian can be written in terms of the on-site spin-${3}/{2}$
operators $\widehat{\vec{S}}_{i}$  as follows \cite{FrKoKoIv11}:
\begin{equation}
  \label{ham-su4-spin}
  \widehat{\mathcal{H}}_{SU(4)} = J\sum_{\langle
    ij\rangle} \Big\{
  -\frac{31}{24}(\widehat{\vec{S}}_{i} \widehat{\vec{S}}_{j})
  +\frac{10}{36}(\widehat{\vec{S}}_{i} \widehat{\vec{S}}_{j})^{2}
  +\frac{2}{9}(\widehat{\vec{S}}_{i} \widehat{\vec{S}}_{j})^{3}
  \Big\},
\end{equation}
with $J=t^{2}/U_{0}$.
For general values of $U_{0}$, $U_{2}$, the effective spin Hamiltonian
still exhibits an enhanced $Sp(4)$ symmetry \cite{WuHuZhang03},
which is special for spin ${3}/{2}$. The effects of symmetry
reduction from $SU(4)$ to $Sp(4)$ were studied in \cite{KV11},
the corresponding perturbation has been shown to be dangerously irrelevant
and is not of interest to us in the present work.
Thus, we consider perturbing the $SU(4)$-invariant model (\ref{ham-su4-spin}) 
\begin{equation}
  \label{pert-spin}
 \widehat{\mathcal{H}}_{SU(4)}  \mapsto \widehat{\mathcal{H}}_{SU(4)} 
 + \lambda \sum_{\langle i j \rangle}(\widehat{\vec{S}}_{i}
 \widehat{\vec{S}}_{j}),\qquad \lambda>0
\end{equation}
by the term that breaks the symmetry down to $SU(2)$ and favors the
antiferromagnetic spin ordering. According
 to the mean-field study
\cite{FrKoKoIv11}, the other sign of $\lambda$ would favor an exotic
phase characterized by the presence of octupolar and quadrupolar
orders and will not be considered here.
Such a perturbation is not possible if only the $s$-wave scattering is
taken into account, but will naturally arise due to the contribution
from the $p$-wave scattering.
Normally, the $p$-wave scattering is 
neglected because the corresponding contributions to the interaction are about a few percent compared
to the $s$-wave ones \cite{Campbell+09}, but in the present case this
is sufficient to break the enhanced symmetry. Beside that,
the effective strength of the $p$-wave scattering can be controlled by quasi-low-dimensional confinement
\cite{GrangerBlume04}.

\section{$SU(2)$-perturbed $CP^{3}$ model: effective theory}

The effective low-energy continuum theory for the $SU(4)$ antiferromagnet
(\ref{ham-su4-spin}) is the well-known $CP^{3}$ model described by the
following euclidean action  \cite{ReadSachdev89,ReadSachdev90}
\begin{equation} 
\label{A-CP3} 
\mathcal{A}_{CP^{3}}=\frac{\Lambda^{d-1}}{2g_{0}}\int
\rd^{d+1}x |\mathcal{D}_{\mu}\vec{z}|^{2} +\mathcal{A}_{\rm top}\,.
\end{equation}
Here, the Planck constant and the lattice spacing are set to unity,
$\Lambda$ is the ultraviolet momentum cutoff, 
the 4-component complex vector field $\vec{z}$ is subjected to the
unit length constraint $\vec{z}^{\dagger}\vec{z}=1$, 
$\mathcal{D}_{\mu}=\partial_{\mu}-\ri A_{\mu}$ is the gauge covariant derivative, and
$A_{\mu}=-\ri(\vec{z}^{\dagger} \partial_{\mu}\vec{z})$ 
is the gauge
field, $x^{0}=c\tau$, $\tau=\ri t$ is the imaginary time.
Assuming for definiteness that the lattice is hypercubic, 
one obtains the limiting velocity $c=2J\sqrt{d}$,  and the
 bare coupling constant  $g_{0}=\sqrt{d}$.  The topological term in the action 
\begin{equation}
\label{Atop}
\mathcal{A}_{\rm top}[\vec{z}]=\int \rd\tau \sum_{j}
\eta_{j}\vec{z}^{\dagger}_{j}\partial_{\tau}\vec{z}_{j}\,,
\end{equation}
where the phase factors $\eta_{j}=\pm1$ take opposite signs at lattice
sites belonging to A and B sublattices,
can be cast in the continuum form only for $d=1$ \cite{Haldane88,ReadSachdev89,ReadSachdev90}.
Without the topological term, the action (\ref{A-CP3}) can be viewed as the energy of the static $(d+1)$ dimensional ``classical'' spin texture.

The 
$CP^{N-1}$
model \cite{Eichenherr78,GoloPerelomov78,DAdda+78,Witten79,ArefevaAzakov80}
has been extensively studied as an effective
theory for $SU(N)$ antiferromagnets
\cite{ReadSachdev89,ReadSachdev90}.
In $d=1$, there is no long-range spin order and  for $N>2$ excitations are gapped at any value of the coupling $g_0$  
even in the presence of the topological term (\ref{Atop}), while in $d=2$ the disordered phase appears if the coupling $g_{0}$
exceeds some $N$-dependent critical value $g_{c}$
\cite{DAdda+78,Witten79}. Numerical work
\cite{Assaad05,KawashimaTanabe07,Beach+09} suggests that for $d=2$
the value $g_{c}N/g_{0}$ lies somewhere between 4 and 5.

The topological term becomes important in the disordered
phase: it drives a
spontaneous breaking of the translational invariance of the underlying
lattice, leading to the twofold degenerate (dimerized) ground state in $d=1$, and in
$d=2$ the ground state degeneracy and the pattern of the resulting ``valence bond
solid'' is lattice-dependent  \cite{Haldane88,ReadSachdev89,ReadSachdev90}.

The leading contribution to the continuum action  from the perturbation (\ref{pert-spin})  is given by the gradient-free term, so the perturbed action takes the form  
\begin{equation}
\label{A-pert}
\mathcal{A}_{AF} =
\frac{\Lambda^{d-1}}{2g_{0}}\int
\rd^{d+1}x \Big\{ |\partial_{\mu}\vec{z}|^{2} -|\vec{z}^{\dagger}\partial_{\mu}\vec{z}|^{2} - m_0^2 \langle \vec{S}\rangle^2\Big\}
 +\mathcal{A}_{\rm top}\,,
\end{equation}
where $m_0^2=2g_0\lambda/(c\Lambda^{d-1}) >0$ is proportional to the perturbation strength,  $ \langle \vec{S}\rangle =\vec{z}^{\dagger}\textrm{S}^{a}\vec{z}$ is the spin average, and $\textrm{S}^{a}$ are spin-${3}/{2}$ matrices, $a=1,2,3$. 
Here, we assume that four components $z_m$ of the complex vector field $\vec{z}$ are directly related to the amplitudes of the four spin-${3}/{2}$ basis states $|\frac{3}{2} m\rangle$, $m=-\frac{3}{2}\ldots\frac{3}{2}$

To analyze the behavior of the perturbed theory, it is convenient to separate the modes becoming massive under the perturbation.  
We parameterize the 4-component field $\vec{z}$ in the following way:
\begin{equation}
\label{z}
z_m=D^{(3/2)}_{mm'}(\alpha,\theta,\varphi) \psi_{m'}(\beta,\vartheta,\phi).
\end{equation}
Here, $\vec{\psi}$   is the spin-${3}/{2}$ state \cite{FrKoKoIv11} taken in the principal axes  of the spin-quadrupolar  tensor 
\begin{equation}
\label{z0}
\psi_{3/2} =
\cos(\beta) \cos\frac{\vartheta}{2}\,, \quad
\psi_{1/2}=\sin\beta \sin\frac{\vartheta}{2} \re^{\ri \phi} ,\quad
\psi_{-1/2}=\sin\beta \cos\frac{\vartheta}{2}\,,\quad
\psi_{-3/2}=\cos\beta \sin\frac{\vartheta}{2} \re^{\ri \phi}\,.
\end{equation}
Such a choice ensures that the tensor $\vec{\psi}^{\dagger} (\textrm{S}^{a}\textrm{S}^{b}+\textrm{S}^{b}\textrm{S}^{a})\vec{\psi}$ is diagonal.
The spin average in the state (\ref{z0}) is given by
\begin{equation}
\label{Sav}
\langle S^z \rangle =\frac{1}{2}\cos\vartheta(4\cos^2\beta -1),\quad \langle S^{+} \rangle =\sin\vartheta\sin\beta(\sqrt{3}\cos\beta \re^{\ri\phi}+\sin\beta \re^{-\ri\phi}).
\end{equation}
The Wigner matrix 
$D^{(j)}$ is the standard $(2j+1)$-dimensional representation of a rotation \cite{LanLif-QM} and depends on three Euler angles $(\alpha,\theta,\varphi)$:
\begin{eqnarray}
\label{D-Wigner}
d_{m'm}^{(j)}(\theta) &=& \left[ \frac{(j+m')!(j-m')!}{(j+m)!(j-m)!} \right]^{1/2} \left(\cos{\frac{\theta}{2}}\right)^{m'+m}\left(\sin{\frac{\theta}{2}}\right)^{m'-m} P^{(m'-m,m'+m)}_{j-m'}(\cos{\theta}),\nonumber\\[2ex]
 D_{m'm}^{(j)}(\alpha, \theta, \varphi)&=&\re^{\ri m' \varphi} d_{m'm}^{(j)}(\theta)  \re^{\ri m \alpha},
\end{eqnarray}
$P^{(a,b)}_{n}(\cos{\theta})$ being the Jacobi polynomials. Rotating the state (\ref{z0}), we obtain the general normalized spin-${3}/{2}$ state characterized by six real parameters
(we omit the overall phase that depends on the gauge). 

Antiferromagnetic perturbation (\ref{pert-spin}) favors field configurations with small $\vartheta$, $\beta$, so it is convenient to introduce three-component real field $\vec{h}=(h_x,h_y,h_z)$:
\begin{equation}
\label{hxyz}
h_x+\ri h_y = \sin\frac{\vartheta}{2} \re^{\ri\phi},\qquad h_z=\sin\beta.
\end{equation}
The perturbation in (\ref{A-pert}) amounts to making $\vec{h}$ massive, as $\langle \vec{S}\rangle^2=\frac{9}{4}-9(h_x^2+h_y^2)- 6h_z^2 + O(h^2)$.

Substituting  the ansatz (\ref{z}), (\ref{hxyz}) into the action (\ref{A-pert}) and retaining up to quadratic terms in powers of $h_{x,y,z}$, one obtains the action in the following form:
\begin{equation}
\label{A-expand}
\mathcal{A}_{\textrm{AF}} = \mathcal{A}_\textrm{NLSM}[\theta,\varphi] + \mathcal{A}_{\mathrm{m}}[\vec{h}] 
+\mathcal {A}_\textrm{int}[\vec{h},\theta,\varphi,\alpha] +\mathcal {A}_\textrm{top}\,,
\end{equation}
where $\mathcal{A}_\textrm{NLSM}$ is the action of the $O(3)$ nonlinear sigma-model,
\begin{equation}
\label{A-NLSM}
\mathcal{A}_\textrm{NLSM}=\frac{3\Lambda^{d-1}}{8g_{0}}\int \rd^{d+1}x \Big\{ (\partial_\mu\theta)^2 +\sin^2\theta(\partial_\mu \varphi)^2\Big\},
\end{equation}
$\mathcal{A}_{\mathrm{m}}$ is the quadratic action of the massive field,
\begin{equation}
\label{A-m}
\mathcal{A}_{\mathrm{m}}=\frac{\Lambda^{d-1}}{2g_{0}}\int \rd^{d+1}x \Big\{ (\partial_\mu \vec{h})^2 +9m_0^2(h_x^2+h_y^2)+6m_0^2h_z^2 \Big\},
\end{equation}
and $\mathcal {A}_\textrm{int}$ describes the interaction, 
\begin{equation}
\label{A-int}
\mathcal {A}_\textrm{int}=\frac{\Lambda^{d-1}}{2g_{0}}\int \rd^{d+1}x \Big\{  
h_z^2 \big[(\partial_\mu\theta)^2 +\sin^2\theta(\partial_\mu \varphi)^2 \big] +(4h_z^2+9h_x^2+9h_y^2)(\partial_\mu\alpha +\cos\theta\partial_\mu\varphi)^2 
+\ldots
\Big\}.
\end{equation}

We integrate out the ``fast'' field $\vec{h}$ and obtain the effective action that depends only on the ``slow'' filelds $\theta$, $\varphi$, $\alpha$:
\begin{equation}
\label{A-SO3}
\mathcal{A}_\textrm{eff} =\frac{\Lambda^{d-1}}{2\Gamma}\int \rd^{d+1}x \Big\{ (\partial_\mu\theta)^2 +\sin^2\theta(\partial_\mu \varphi)^2\Big\}
+\frac{\Lambda^{d-1}}{2G}\int \rd^{d+1}x (\partial_\mu\alpha +\cos\theta\partial_\mu\varphi)^2 +\mathcal {A}_\textrm{itop}\,,
\end{equation}
where the renormalized couplings $\Gamma$, $G$ are determined  by the equations
\begin{eqnarray}
\label{G}
\frac{\Lambda^{d-1}}{\Gamma} = \frac{\Lambda^{d-1}}{\Gamma_0} +\frac{1}{(2\piup)^{d+1}}\int_{k<\Lambda} \frac{\rd^{d+1}k}{k^2+6m_0^2}\,, \qquad \Gamma_0=\frac{4g_0}{3}\,,\nonumber\\
\frac{\Lambda^{d-1}}{G}=\frac{1}{(2\piup)^{d+1}}\int_{k<\Lambda} \rd^{d+1}k\left\{ \frac{4}{k^2+6m_0^2}+\frac{18}{k^2+9m_0^2}\right\}.
\end{eqnarray}
Beside terms of higher than quadratic order in $\vec{h}$,  in the interaction (\ref{A-int})  we have omitted several types of terms that will not contribute to the renormalized action at the one-loop level. Namely, terms proportional to  $h_{z}h_{x,y} (\partial_\mu \Phi)^2$, $h_z(\partial_\mu \Phi)^2$, where $\Phi$ denotes any of the slow fields, would generate terms of 
the fourth and higher order in gradients of $\Phi$.  The other omitted term, of the structure $(h_x\partial_\mu h_y - h_y\partial_\mu h_x) (\partial_\mu \Phi)$, yield  contributions that vanish   after integration over the wave vector.

For small AF perturbations, $m_0/\Lambda\ll1$, from (\ref{G}) one has 
\begin{eqnarray}
\label{G-d1-d2}
 \Gamma&\simeq& \frac{2\piup}{\ln\frac{\Lambda}{m_0}}\,, \qquad G\simeq \frac{\piup}{11\ln\frac{\Lambda}{m_0}}\quad
\text{for $d=1$}, \nonumber\\
 \Gamma&\simeq&\frac{\Gamma_0}{1+\Gamma_0/2\piup^2} +O(m_0/\Lambda), \qquad G\simeq \frac{11}{\piup^2} +O(m_0/\Lambda)\quad \text{for $d=2$}.
\end{eqnarray}

The action (\ref{A-SO3}) describes the $O(3)\times O(2)$ NLSM that is encountered as an effective theory of frustrated antiferromagnets \cite{DombreRead89,Azaria+92}. It can be recast in the form 
\begin{equation}
\mathcal{A}_\textrm{eff}=\frac{\Lambda^{d-1}}{2\Gamma} \int \rd^{d+1}x  \text{Tr}\left(\partial_\mu R^T P \partial_\mu R\right),\qquad P=\text{diag}(1,1,\zeta),\qquad \zeta=\Gamma/G,
\end{equation}
where the matrix field $R\in SO(3)$ is the rotation matrix, and filelds $\theta$, $\varphi$, $\alpha$ are connected to $R$ via the standard relations
\begin{equation}
\label{dR}
\Omega_\mu = 
\begin{pmatrix} 
0 & -\omega_{\mu 3} & \omega_{\mu 2} \\ 
\omega_{\mu 3} & 0 & -\omega_{\mu 1} \\
-\omega_{\mu 2} & \omega_{\mu 1} & 0
\end{pmatrix} = \partial_\mu R  R^{\mathrm{T}},
\end{equation}
where $\omega_{\mu a}$ are the rotation ``frequencies'' in the rest frame
\begin{equation}
\label{omega}
\omega_{\mu 1}=-\sin\alpha\partial_\mu\theta +\cos\alpha\sin\theta\partial_\mu\varphi, \quad 
\omega_{\mu 2}=\cos\alpha\partial_\mu\theta +\sin\alpha\sin\theta\partial_\mu\varphi, \quad 
\omega_{\mu 3}=\partial_\mu\alpha +\cos\theta\partial_\mu\varphi . 
\end{equation}
The field $R$ may be visualized as rotating axisymmetric top. In the standard $O(3)$ NLSM, $\zeta=0$ and this top is an ``arrow'' (a unit vector defined by the polar and azimuthal angles $\theta$, $\varphi$) with one inertia momentum equal to zero.  
One can see that fluctuations of massive fields lead to a dynamic generation of the third inertia momentum, so the effective theory of the AF-perturbed model is not the standard  $O(3)$ NLSM as one might naively guess, but rather the $O(3)\times O(2)$ NLSM. 
Properties of the latter model are well known \cite{Azaria+92,Azaria+92a,ApelWintelEverts92}. In one dimension, $\zeta$ flows to the $O(4)$ fixed point $\zeta=1$, while $\Gamma$ flows to infinity, indicating the dynamic generation of a finite correlation length. For $d=2$, the  the $O(3)\times O(2)$ NLSM has long-range AF order, couplings $\Gamma$, $\zeta$ get renormalized but stay finite. 
We have checked that the similar dynamic generation of the third inertia momentum occurs for spin-1 systems close to the antiferromagnetic $SU(3)$ point, 
resulting in $O(3)\times O(2)$ NLSM as the effective model, so one may expect this result to be valid for general $S$.

\section{Multiplet binding of topological excitations}

Consider the fate of topologically nontrivial excitations of the $SU(4)$ -symmetric model (\ref{A-CP3}) under AF perturbation (\ref{A-pert}) breaking the symmetry down to $SU(2)$. In $(1+1)$ dimensions, such excitations (``skyrmions'') in $CP^{N-1}$ model are characterized by the nonzero   
topological
charge \cite{DAdda+78} that is essentially the winding number of the overall phase taken over a contour at infinity,
\begin{equation} 
\label{q-CPN} 
q_{CP^{N-1}}=-\frac{1}{2\piup} \oint \vec{A}\cdot \rd\vec{l} =   -\frac{1}{2\piup} \int \rd^{2}x \epsilon_{\mu\nu}(\partial_{\mu}A_\nu)=  -\frac{\ri}{2\piup}\int \rd^{2}x \epsilon_{\mu\nu}(\partial_{\mu}\vec{z}^{\dagger}
\partial_{\nu}\vec{z})
\end{equation}
and for $d=1$ this charge is directly related to the topological term in the action (\ref{A-CP3}), $\mathcal{A}_\textrm{top}= \ri\piup q_{CP(3)}$. 
The topological charge density is proportional to the dimerization order parameter; the ground state of the $SU(4)$ -symmetric model (\ref{A-CP3}) has finite topological charge density and thus is spontaneously dimerized~\cite{ReadSachdev90}.

The effect of the $SU(4)\mapsto SU(2)$ perturbation
on the topological charge can be illustrated  by the following simple observation: finite $\lambda$ favors field configurations with the maximum spin length, i.e., with $\vec{h}=0$.
Such field configurations are given by
\begin{equation}
\label{zAF}
z_m=D^{(3/2)}_{mm'}(\alpha,\theta,\varphi) \psi^{(0)}_{m'}\,,\qquad \psi^{(0)}_m=\delta_{3/2,m} \,.
\end{equation}
Substituting the above ansatz into (\ref{q-CPN}), 
one straightforwardly obtains
\begin{eqnarray} 
\label{q=3Q}
q_{CP^3}&=&\frac{3}{4\piup}\int \rd^{2}x\sin\theta
\epsilon_{\mu\nu}(\partial_{\mu}\theta)( \partial_{\nu}\varphi) = 3Q_{O(3)}\,,
\end{eqnarray}
where the topological charge $Q_{O(3)}$ is the winding number of the $S^{2}\mapsto
S^{2}$ mapping characterizing the space-time distribution of the unit vector
$\vec{n}(\theta,\varphi)$. 
It should be remarked that although the homotopy group $\pi_2(SO(3))=0$, one can still define the $O(3)$-NLSM topological charge of the spin texture $Q_{O(3)}$
in the usual way via the unit vector $\vec{n}(\theta,\varphi)$ that corresponds to the local direction of the N\'eel vector. 
The AF perturbation thus favors
$\vec{z}$-field configurations with charge $q_{CP^3}$  being a multiple of $3$. 
One may conclude that unit-charge skyrmions of  the $CP^3$ model bind into \emph{triplets} under the influence of the AF perturbation. Such a triplet is the well-known unit-charge skyrmion  (Belavin-Polyakov soliton \cite{BelavinPolyakov75}) of the $O(3)$ NLSM.

This is completely analogous to the formerly noted effect\cite{IvKhymKol08,Kolezhuk08} of topological binding of skyrmions into \emph{pairs} in the $SU(3)$ spin-1 antiferromagnet under the AF perturbation lowering the symmetry from $SU(3)$ to $SU(2)$.

This statement is easily generalized: consider the $SU(2S+1)$ antiferromagnet with the underlying spin $S$. Assume that the enhanced symmetry is broken down to $SU(2)$ by the perturbation that favors the field configuration with the maximal spin length, as in (\ref{A-pert}):
\begin{equation}
\label{z-S}
\vec{z}=D^{(S)}(\alpha,\theta,\varphi) \vec{\psi}^{(0)}
= \re^{\ri \varphi \hat{S}_3}\re^{\ri \theta \hat{S}_2}\re^{\ri \alpha \hat{S}_3}\vec{\psi}^{(0)}, \qquad \psi^{(0)}_{m}=\delta_{S,m}\,.
\end{equation}
Then,  
\begin{eqnarray}
\label{z-der}
   \partial_{\mu} \vec{z} &=& \ri (\partial_{\mu} \varphi) \hat{S}_3 \vec{z} 
+ \ri (\partial_{\mu}\theta)  \re^{\ri \varphi \hat{S}_3} \hat{S}_2 \re^{\ri \theta \hat{S}_2}\re^{\ri \alpha \hat{S}_3} \vec{\psi}^{(0)}
+\ri (\partial_{\mu} \alpha) S \vec{z}\,, \nonumber\\
\epsilon_{\mu\nu}   \partial_{\mu} \vec{z}^\dagger  \partial_{\nu}  \vec{z} &=& \varepsilon_{\mu \nu} (\partial_{\mu} \theta) (\partial_{\nu} \varphi ) \left(\vec{\psi}^{(0)}\right)^{\dagger} \re^{-\ri \theta \hat{S}_2} ( \hat{S}_2 \hat{S}_3 - \hat{S}_3 \hat{S}_2) \re^{\ri \theta \hat{S}_2}  \vec{\psi}^{(0)} \nonumber\\ 
&=& \ri S \sin{\theta} \varepsilon_{\mu \nu} (\partial_{\mu} \theta)( \partial_{\nu} \varphi).
\end{eqnarray}
Substituting this into (\ref{q-CPN}), we see that the $CP^{2S}$ charge takes the form 
\begin{equation}
\label{q=2SQ}
    q_{CP^{2S}}= 2S Q_{O(3)}\,.
\end{equation}
Thus, in  $SU(2S+1)$ antiferromagnets with underlying spin $S$, AF perturbation that breaks the enhanced symmetry down to $SU(2)$ leads to the binding of unit-charge skyrmions of  the $CP^{2S}$ model to $2S$-\emph{multiplets}.

Strictly speaking, in $(1+1)$ dimensions, skyrmions considered above, are not excitations, but instanton events.
The same effect obviously holds for ``monopoles'' in $(2+1)$ dimensions (instanton events changing the skyrmion topological quantum number $q_{CP^{2S}}$).

For the $(2+1)$ dimensional case, skyrmions may be viewed as static solitons in two spatial dimensions, and
similarly, in $d=3$ monopoles they may be viewed as static solitons (``hedgehogs'');  the same reasoning on multiplet binding applies. 

To show that explicitly, one may look at the ``skyrmion current''  $\vec{j}=(2\piup)^{-1} \nabla \times \vec{A}$,
whose flux through a closed surface surrounding the monopole, $\widetilde{q}=\oint \vec{j}\cdot \rd\vec{S} $,  determines the monopole charge $\widetilde{q}$. 
A calculation essentially following (\ref{z-der}) shows that
\begin{equation} 
\label{current-CPN} 
j_{a}=-\frac{\ri}{2\piup}
  \varepsilon_{abc}(\partial_{b}\vec{z}^{\dagger} \partial_{c}\vec{z}) 
= \frac{S}{2\piup}  \varepsilon_{abc} \sin\theta(\partial_{b} \theta)( \partial_{c} \varphi) =2S J_a\,,
\end{equation}
where $\vec{J}$ is  the corresponding skyrmion current of the $O(3)$ NLSM. Thus, $\widetilde{q}=\pm1$ monopoles bind into $2S$ multiplets under the influence of the perturbation.

The topological term in the action in the $(2+1)$ dimensional case 
is determined by monopole events and can not be expressed in the continuum limit as it is lattice-dependent \cite{Haldane88,ReadSachdev90}. On the square lattice, it is given by
\begin{equation} 
\label{AB-2d} 
\mathcal{A}_\textrm{top}=\frac{\ri}{2}\piup n_{\mathrm{c}}\sum_{\vec{r}_{i}} \zeta(\vec{r}_{i}) 
\widetilde{q}_{i}\,,
\end{equation}
where the sum is over the locations $\vec{r}_{i}$ of monopoles having the charge
$ \widetilde{q}_{i}$, and factors $\zeta(\vec{r}_{i})$ take on values $0$, $1$,
$2$, $3$ for $\vec{r}_{i}$ belonging to the four dual sublattices $W$, $X$, $Y$,
$Z$ respectively (see figure~7 of \cite{ReadSachdev90}), and $n_{\mathrm{c}}$ is the ``colour number'' that in our case is equal to 1.  

In $(2+1)$ dimensions, the ground state of the $CP^{N-1}$ model can have the long-range order or can be disordered, depending on the value of the coupling $g_0$.
In the ordered phase, the topological term is ineffective. However, if for some reason the coupling gets driven over the critical value (e.g., due to the presence of next-nearest neighbor  interactions) and we land in the disordered phase, the topological term becomes important: it  leads to the ground state with
nonzero monopole density and thus to the spontaneous dimerization (i.e., breaking of the translation
symmetry) \cite{ReadSachdev90}. The dimerization pattern of the ground state depends on the value of $n_{\mathrm{c}}$: it is twofold degenerate
for $n_{\mathrm{c}}=2 \bmod 4$, fourfold degenerate for $n_{\mathrm{c}}=1 \bmod 4$ or $n_{\mathrm{c}}=3 \bmod 4$, and non-degenerate (with unbroken translational invariance) for $n_{\mathrm{c}}=0 \bmod 4$. Thus, when the $SU(N)$-symmetric antiferromagnet gets perturbed as in (\ref{A-pert}) by the $SU(2)$ term favoring the N\'eel order, $2S$-multipletting of unit-charge monopoles leads to $\widetilde{q}_i$ in (\ref{AB-2d}) multiplied by $2S$, which is equivalent to changing the number of ``colours'' $n_{\mathrm{c}}$ from 1 to $2S$. 
The ground state becomes respectively twofold degenerate for odd-integer spins $S=2n+1$, stays fourfold degenerate for half-integer $S$, and  is non-degenerate for even-ineger  $S=2n$. This result coincides with the conclusion obtained by Haldane \cite{Haldane88} in the framework of the $O(3)$ NLSM analysis.

\section{Summary}

 We have considered the 
low-dimensional  spin-$S$ antiferromagnet 
on a bipartite lattice, close to the point with the enhanced $SU(2S+1)$ symmetry which is decribed in the continuum field approximation by the $CP^{2S}$ model,
and studied the consequences of explicit symmetry breaking from $SU(2S+1)$ to $SU(2)$ (with the appropriate sign that favors the N\`eel order). 
This model is motivated by the
physics of cold spinor bosonic atoms in optical lattices, and the symmetry-breaking perturbation can be associated with weak interactions such as $p$-wave scattering that are usually neglected. 
We derive the effective theory for the perturbed system, and show that it is not the standard  $O(3)$ nonlinear sigma model (NLSM) as one might naively guess, but rather the $O(3)\times O(2)$ NLSM. This occurs due to the dynamic generation of the third inertia momentum for the ``spin arrow'', caused by fluctuations of massive fields that correspond to non-axisymmetric deformations of the quadrupolar tensor.
We have further shown that under the influence of the $SU(2S+1)\mapsto SU(2)$ perturbation unit-charge topological excitations (skyrmions and monopoles) of the $CP^{2S}$ model bind into $2S$ multiplets that correspond to excitations with unit $O(3)$ topological charge defined in terms of the unit N\`eel vector.



%
%

\ukrainianpart

\title{Порушення симетрії $SU(N)\to SU(2)$ в квантових антиферомагнетиках}
\author{О.К. Колежук \refaddr{KNU,IMAG}, Т.Л. Завертаний\refaddr{KNU}}
\addresses{
\addr{KNU} Інститут високих технологій, Київський національний університет імені Тараса Шевченка, \\ МОН України, Київ 03022, Україна
\addr{IMAG} Інститут магнетизму НАН України та МОН України, просп. Вернадського 36--Б, Київ 03142, Україна
}

\makeukrtitle

\begin{abstract}
Ми досліджуємо $SU(2)$-симетричну систему спіну ${3}/{2}$ на двороздільній ґратці, поблизу антиферомагнітної $SU(4)$-симетричної точки, що може бути описана 
$CP^{3}$ моделлю зі збуренням, що порушує симетрію до $SU(2)$ і заохочує неєлівське впорядкування.  Показано, що ефективною теорією для збуреної моделі є не звичайна $O(3)$ нелінійна сігма-модель (НЛСМ), 
а $O(3)\times O(2)$ НЛСМ.  Разом з тим, топологічний заряд  $Q$ 
$O(3)$-НЛСМ типу для спінової текстури одиничного вектора антиферомагнетизму може бути введений звичайним чином.
Показано, що в присутності збурення топологічний заряд $q$ поля $CP^{3}$ пов'язаний з зарядом  $Q$ 
$O(3)$-НЛСМ типу (для спінової текстури одиничного вектора антиферомагнетизму) співвідношенням 
$3Q=q$, тому під дією збурення скірміони моделі $CP^{3}$ з одиничним зарядом зв'язуються в триплети.
Також показано, що у загальному випадку спіну $S$, порушення симетрії з
$SU(2S+1)$ до $SU(2)$ приводить до співвідношення $2S Q_{O(3)}=q_{CP^{2S}}$ між топологічними зарядами $CP^{2S}$ та $O(3)$ типу, тобто можна очікувати що скірміони будуть у цьому випадку зв'язуватися в $2S$-мультиплети.
\keywords фрустровані магнетики, скірміони, холодні гази в оптичних ґратках
\end{abstract}

\end{document}